\documentclass[10pt,leqno]{amsart}
\usepackage[T1]{fontenc}
\usepackage{mathtools}
\usepackage{amssymb,amsthm,amsmath}
\usepackage{yhmath}
\usepackage{xcolor,paralist,hyperref,fancyhdr,etoolbox}
\newtheorem{theorem}{Theorem}[]
\newtheorem{definition}[theorem]{Definition}
\newtheorem{example}[theorem]{Example}
\newtheorem{lemma}[theorem]{Lemma}
\newtheorem{claim}[theorem]{Claim}

\begin{document}
\title{A precise proof of the $n$-variable Beki\v{c} priniciple } 
\author[J Xu]{Xu Jun}
\date{\today}
\address{}
\email{willxujun@hotmail.com}

\begin{abstract}
We provide a proof of the $n$-ary Beki\v{c} principle, which states that a vectorial fixpoint of size $n$ can be written in terms of nested fixpoints in each coordinate according to lexicographic order. The proof is inductive. 
\end{abstract} 
\maketitle

\bigskip

The Beki\v{c} principle provides a way to express vectorial fixed points of monotonic functions in terms of iterated fixpoints of the function's components. While standard texts often showcase the 2-variable case, the $n$-ary case for $n>3$ has been noticeably missing from mathematics literature, yet it should be a direct generalisation of the $n=2$ case. In this paper we provide an inductive proof of the $n>3$ case. In light of the recent need to formalize in a theorem prover the general $n$-ary case of this principle, as used in \cite{Abou_El_Wafa_2024}, the proof must not depend on human intuition and must follow rigorous logic. This in particular forbids a proof sketch that writes an explicit expression for an unknown $n$ (think of a typical list reversal statement \verb|rev rev ls = ls|. This cannot be proven in a theorem prover by saying ``suppose $ls = [a_1,\dots,a_n]$...'' but one must apply rigorous structural induction on $ls$). 

As a reminder, we first state the $n=2$ version for least upper bounds.

\begin{theorem}[Beki\v{c}]
    Let $E,F$ be a complete lattices, and let $f: E\times F \rightarrow E\times F$ be a monotone function. Let $f_1,f_2$ be the projection of $f$ onto the respective coordinates. Then the following equality holds:
    \[\mu \begin{bmatrix}
        x \\ y
    \end{bmatrix}. \begin{bmatrix}
        f_1(x,y) \\
        f_2(x,y)
    \end{bmatrix} = \begin{bmatrix}
        \mu x. f_1(x,\mu y.f_2(x,y)) \\
        \mu y. f_2(\mu x.f_1(x,y),y))
    \end{bmatrix}\]
\end{theorem}
\begin{proof}
    See for example Lemma 1.4.2 of \cite{arnold2001rudiments}.
\end{proof}

The equality only asserts that RHS is one way to express LHS. It does not assert that it is the only way. In fact, the question of whether there are shorter ways to express LHS (less than exponentially long) for $n>2$ is open.

One might venture a guess for what higher dimensions look like. For example, for $n=3$,
\begin{equation}
\label{eqn_3}
    \mu \begin{bmatrix}
        x \\ y \\ z
    \end{bmatrix}. \begin{bmatrix}
        f_1(x,y,z) \\
        f_2(x,y,z) \\
        f_3(x,y,z)
    \end{bmatrix} = \begin{bmatrix}
        \mu x. f_1(x,\mu y.f_2(x,y,\mu z. f_3(x,y,z)), \mu z. f_3(x,\mu y.f_2(x,y,z),z)) \\
        \mu y. f_2(\mu x.f_1(x,y,\mu z. f_3(x,y,z)), y, \mu z. f_3(\mu x. f_1(x,y,z), y ,z)) \\
        \mu z. f_3(\mu x.f_1(x,\mu y. f_2(x,y,z),z),\mu y.f_2(\mu x.f_1(x,y,z),y,z),z)
    \end{bmatrix}
\end{equation}

Observing the nesting structure of the above, one can naturally organise the fixpoints into a tree. Such tree corresponds to strings of length $n$ over the alphabet $\{1,\dots,n\}$, without repetition, according to lexicographic order. It turns out this is the correct intuition; next we make it precise and state the $n$-variable Beki\v{c} principle.

 For general $n$, we use a recursive definition for nested fixpoints. We fix some notations. Write $nested$ as the recursive, nested fixpoint we want to define. Write $\lambda j. [V(j)]$ as a vector in $L_1\times\dots\times L_n$ such that the $j$-th coordinate is given by $V(j)\in L_j$. Given a binary predicate $b$ and values $G,H \in L_i$, write $?b:G;H$ as the value in $L_i$ given by if-then-else: if $b$ evaluates to True, then $?b:G;H = G$; else $?b:G;H = H$. In particular, we write $v\neq undef$ as the predicate saying that $v$ is NOT undefined. We use a special symbol $undef$ to indicate an undefined value. For a map $B: \{1\dots n\} \rightarrow (\bigsqcup L_i)\sqcup \{undef\}$, we write $B(j\coloneq v)$ as the modified map that sends $j$ to $v$ and $i$ to $B(i)$ for $i\neq j$.

\begin{definition}
    Let $L_1,\dots,L_n$ be complete lattices. Let $f:L_1\times \dots \times L_n\rightarrow L_1\times\dots\times L_n$ be a monotone function. Let $f_i$ be the projection of $f$ to the $i$-th coordinate. Fix a map $B:\{1\dots n\} \rightarrow (\bigsqcup L_i)\sqcup \{undef\}$ mapping coordinate positions $i$ to $L_i$. We define
    \begin{equation}
    \label{eqn_nested_fp}
        nested(n,i,B,f) = \mu x_i.f_i(\lambda j. [?j=i:x_i;?(B(j)\neq undef): B(j); nested(n,j,B(i\coloneq x_i), f)])
    \end{equation}
    A nested fixpoint for $f:L_1\times \dots \times L_n\rightarrow L_1\times\dots\times L_n$, of size $n$, at coordinate $i$, is given by
    \[ nested(n,i,B_0,f) \]
    where $B_0(i)=undef$ for each $i\in \{1\dots n\}$.
\end{definition}

To see that the above procedure (\ref{eqn_nested_fp}) terminates, consider induction on the unbound coordinates of $B$. If $B$ binds all coordinates, then the definition has no recursion thus terminates. If $B$ has $x_1,\dots,x_k$ unbound, then the if-then-else clause in the definition tests for an unbound $j$-th coordinate for $B$. It goes into recursion only when $B$ does not bound the $j$-th coordinate, and the recursive $nested$ clause uses $B(j\coloneq x_j)$ which has one fewer unbound coordinate than $B$. Hence the recursive clause terminates by induction, hence the whole definition eventually terminates.

Intuitively, the recursive case produces $f_i$ applied to a vector that is bound to $x_i$ in the $i$-th coordinate and recursively go down one level, with $i$-th coordinate bound to $x_i$ thereafter. The information for bound values is recorded in $B$.

\begin{example}
    We expand $nested(3,B_0,f)$ for illustration. Firstly we reduce to:
    \[\mu x_1.f_1(\lambda j. [?j=1:x_1;?(B_0(j)\neq undef): B_0(j); nested(3,B_0(1\coloneq x_1), f)]\]
    The vector $\lambda j. [?j=1:x_1;?(B_0(j)\neq undef): B_0(j); nested(3,j,B_0(1\coloneq x_1), f)]$ can be written as
    \[(x_1, nested(3,2,B_0(1\coloneq x_1),f), nested(3,3,B_0(1\coloneq x_1),f))\]
    We expand the depth 2 nested fixed points:
    \begin{align*}
        nested(3,2,B_0(1\coloneq x_1),f) = \mu x_2.f_2(x_1,x_2, nested(3,3,B_0(1\coloneq x_1,2\coloneq x_2),f))\\
        nested(3,3,B_0(1\coloneq x_1),f) = \mu x_3.f_3(x_1, nested(3,2,B_0(1\coloneq x_1,3\coloneq x_3),f),x_3)
    \end{align*}
    We expand the depth 3 nested fixed points. Note we are in the base case:
    \begin{align*}
        nested(3,3,B_0(1\coloneq x_1,2\coloneq x_2),f) = \mu x_3.f_3(x_1,x_2,x_3) \\
        nested(3,2,B_0(1\coloneq x_1,3\coloneq x_3),f) = \mu x_2.f_2(x_1,x_2,x_3)
    \end{align*}
    Substituting into the original fixpoint, we get that $nested(3,1,B_0(1\coloneq x_1),f)$ equals:
    \begin{equation*}
        \mu x_1.f_1(x_1,\mu x_2.f_2(x_1,x_2,\mu x_3.f_3(x_1,x_2,x_3)), \mu x_3.f_3(x_1,\mu x_2.f_2(x_1,x_2,x_3),x_3))
    \end{equation*}
    This is exactly the first coordinate of equation \ref{eqn_3}.
\end{example}

Now we are ready to state the $n$-variable Beki\v{c} principle. Write $\mu\vec{x}.f(\vec x)$ for \[\mu[x_1,\dots,x_n]^T.[f_1(x_1,\dots,x_n),\dots,f_n(x_1,\dots,x_n)]^T\]
for some $n$ that is implicitly understood.

\begin{theorem}[$n$-ary Beki\v{c}]
\label{thm_bekic}
    For any $i=1,\dots,n$, the $i$-th coordinate of $\mu \vec x.f$ is equal to
    \[nested (n,i,B_0,f)\]
    where $B_0(j) = undef$ for all $j=1,\dots,n$.
\end{theorem}

To prove the theorem we need the notion of specializing a function to a value. Given a monotone function $f:L_1\times \dots\times L_n\rightarrow L_1,\dots,L_n$ and some $a_i\in L_i$, we call the function 
\[(x_1,\dots,\widehat{x_i},\dots,x_n)\mapsto (f_1(x_1,\dots,a_i,\dots,x_n),\dots,\widehat{f_i(x_1,\dots,a_i,\dots,x_n)},\dots,f_n(x_1,\dots,a_i,\dots,x_n))\]
the \textbf{specialization} of $f$ at $a_i$, written as $Sp_{i,a_i}f$. Note that $Sp_{i,a_i}f$ has domain and codomain $L_1\times\dots \times\widehat{L_i}\times \dots \times L_n$. For $(y_1,\dots,y_{n-1})$, we define the $(i,a_i)$-shifted vector to be
\begin{equation}
    \label{eqn_vec_shift}
    E_{i,a_i}\vec y = \lambda j\in\{1\dots n\}.?j<i;y_j;j=i:a_i; y_{j-1}
\end{equation}
or more intuitively $(y_1,\dots,a_i,y_i,\dots,y_{n-1})$.

Note there is a coordinate vector shift in the projections of specialized function. Let $j\in\{1,\dots,n-1\}$. Then:
\begin{equation}
\label{eqn_shift}
    (Sp_{i,a_i}f)_j(y_1,\dots,y_{n-1}) = \begin{cases}
        f_j(E_{i,a_i}\vec y) & \text{if } 1\leq j<i \\
        f_{j+1}(E_{i,a_i}\vec y) & \text{if } i\leq j \leq n-1
    \end{cases}
\end{equation}

There is a corresponding shift in $B$. Given $j\in\{1,\dots,n-1\}$:
\begin{equation}
\label{eqn_shiftB}
    (Sp_iB)(j) = \begin{cases}
        B(j) & \text{if } 1\leq j < i\\
        B(j+1) & \text{if } i\leq j \leq n-1
    \end{cases}
\end{equation}

What variables do $Sp_{i,a_i}f$ use? When we construct a nested fixpoint
$nested(n-1,k,B,Sp_{i,a_i}f)$, the algorithm is going to create a clause
$\mu x_k.(Sp_{i,a_i}f)_k$, which binds the $x_k$ position of $Sp_{i,a_i}f$ and projects to coordinate $k$. The ambiguity with this expression is that often we would make $\mu x_k. f_k(\vec v)$ with some special vector $\vec v$ that fixes one coordinate of $f_k$, making it equal to some $Sp_{j,a_j}f$. Then $k\in\{1,\dots,n\}$ has to be put to the appropriate coordinate in the $(n-1)$-ary input vector, and  $(Sp_{i,a_i}f)(x_1,\dots,\widehat{x_i},\dots,x_n)$ has input vector $y_1,\dots,y_{n-1}$, so there is a skip of $i$ when mapping to $f$'s coordinates. We keep to the convention that any $n$-ary function has input vector coordinates numbered from 1 to $n$, and track any transformation between $(n-1)$-ary and $n$-ary input vectors with a map, similar to the simplicial category.

\begin{lemma}
\label{lem_shift}
    Let $i,j\in \{1,\dots,n\}$, where $n>1$. For $i<j$, we have
    \[ nested(n,i,B(j\coloneq a_j), f) = nested(n-1,i,Sp_jB,Sp_{j,a_j}f) \]
    For $i>j$, we have
    \[ nested(n,i,B(j\coloneq a_j), f) = nested(n-1,i-1,Sp_jB,Sp_{j,a_j}f) \]
\end{lemma}
\begin{proof}
By induction on the number of unbound coordinate in $B$. We assume $B$ does not bind $j$ so that $B(j\coloneq a_j)$ is a genuine modification. Hence $B$ has at least 1 unbound coordinate. The base case says $B$ only has 1 unbound coordinate and that has to be $j$. Then LHS reduces to $\mu x_i.f_i(B(1),\dots,a_j,\dots,B(n))=f_i(B(1),\dots,a_j,\dots,B(n))$. Meanwhile RHS has $B$ binding every coordinate $1,\dots,\hat j,\dots,n$, as $Sp_{j,a_j}f$ acts on $L_1\times\dots\times \widehat L_j\times \dots\times L_n$. When $i<j$, RHS reduces to $\mu x_i.(Sp_{j,a_j}f)_i(B(1),\dots,\widehat {B(j)},\dots,B(n)) = \mu x_i.f_i(B(1),\dots,a_j,\dots,B(n))=f_i(B(1),\dots,a_j,\dots,B(n))$. When $i>j$, RHS takes the value $\mu y_{i-1}.(Sp_{j,a_j}f)_{i-1}(B(1),\dots,\widehat{B(j)},\dots,B(n)) = \mu y_{i-1}.f_i(B(1),\dots,a_j,\dots,B(n))=f_i(B(1),\dots,a_j,\dots,B(n))$. So the base case is proven.

For the inductive case, consider $i<j$. Unwrap the definition of LHS to get 
\[\mu x_i.f_i(\lambda k. ?k=i:x_i; ?k=j:a_j; ?B(k)\neq undef: B(k); nested(n,k,B(j\coloneq a_j,i\coloneq x_i),f))\]
Since $B(i\coloneq x_i)$ has 1 fewer variable bound than $B$, IH can be applied to yield
\[nested(n,k,B(j\coloneq a_j,i\coloneq x_i),f) = \begin{cases}
    nested(n-1,k,Sp_j(B(i\coloneq x_i)), Sp_{j,a_j}f)  & \text{if } k<j\\
    nested(n-1,k-1,Sp_j(B(i\coloneq x_i)), Sp_{j,a_j}f)  & \text{if } k>j
\end{cases}\]
Note that $Sp_j(B(i\coloneq x_i))=Sp_jB(i\coloneq x_i)$ when $i<j$. Hence the expression 
\[f_i(\lambda k. ?k=i:x_i; ?k=j:a_j; ?B(k)\neq undef: B(k); nested(n,k,B(j\coloneq a_j,i\coloneq x_i),f))\]
with a map of coordinates $h\in\{1,\dots,n-1\}$ to $k\in\{1,\dots n\}$ skipping $j$, evaluates to
\begin{multline*}
 (Sp_{j,a_j}f)_i(\lambda h\in\{1\dots n-1\}. ?h=i:x_i;?Sp_jB(h)\neq undef: Sp_jB(h);\\
 ?h < j: nested(n-1,h,Sp_jB(i\coloneq x_i), Sp_{j,a_j}f); \\
 nested(n-1,h+1-1,Sp_jB(i\coloneq x_i),Sp_{j,a_j}f) \\
\end{multline*}
which is precisely
\begin{multline*}
 (Sp_{j,a_j}f)_i(\lambda h\in\{1\dots n-1\}. ?h=i:x_i;?Sp_jB(h)\neq undef: Sp_jB(h);\\
 nested(n-1,h,Sp_jB(i\coloneq x_i), Sp_{j,a_j}f))
\end{multline*}
Adding $\mu x_i$ at the front we get equality to RHS.

When $i>j$, unwrap the LHS to get the same expression. We have a symmetric IH, so it can be applied in the same manner to yield
\[nested(n,k,B(j\coloneq a_j,i\coloneq x_i),f) = \begin{cases}
    nested(n-1,k,Sp_j(B(i\coloneq x_i)), Sp_{j,a_j}f)  & \text{if } k<j\\
    nested(n-1,k-1,Sp_j(B(i\coloneq x_i)), Sp_{j,a_j}f)  & \text{if } k>j
\end{cases}\]
However, this time $Sp_j(B(i\coloneq x_i))= (Sp_jB)(i-1\coloneq x_i)$. We evaluate to
\begin{multline*}
 (Sp_{j,a_j}f)_{i-1}(\lambda h\in\{1\dots n-1\}. ?h=i-1:x_i;?Sp_jB(h)\neq undef: Sp_jB(h);\\
 ?h < j: nested(n-1,h,Sp_jB(i-1\coloneq x_i), Sp_{j,a_j}f); \\
 nested(n-1,h+1-1,Sp_jB(i-1\coloneq x_i),Sp_{j,a_j}f) \\
\end{multline*}
which is
\begin{multline*}
 (Sp_{j,a_j}f)_{i-1}(\lambda h\in\{1\dots n-1\}. ?h=i-1:x_i;?Sp_jB(h)\neq undef: Sp_jB(h);\\
 nested(n-1,h,Sp_jB(i-1\coloneq x_i), Sp_{j,a_j}f))
\end{multline*}
We observe that the $\mu x_i$ in front actually corresponds to the $i-1$-th coordinate of the input vector of $Sp_{j,a_j}f$. Hence we get precisely
\[nested(n-1,i-1,Sp_jB,Sp_{j,a_j}f)\]
which is RHS.
\end{proof}

We are ready to prove the $n$-ary Beki\v{c} principle.

\subsection{Proof of theorem \ref{thm_bekic}}
Write the vectorial fixpoint as
\[\mu \vec x.f = \begin{bmatrix}
    a_1 \\ a_2 \\ \dots \\ a_n
\end{bmatrix}\]
We prove two inequalities separately. First we make a claim about $a_i$ being larger. We use $[i_1\coloneq a_{i_1},\dots,i_k\coloneq a_{i_k}]$ to denote the map $B_0(i\coloneq a_i,\dots,i_k\coloneq a_{i_k})$, where $B_0$ is our favorite empty map undefined on every coordinate.
\begin{claim}
\label{claim_ailarge}
    Let $a_1,\dots,a_n$ be components of the vectorial fixpoint of a monotone function $f$. Then $a_i\geq f_i(\lambda j. ?j=i:a_i; nested(n,j,[i\coloneq a_i],f)$.
\end{claim}
\begin{proof}
    By induction on $n$. Let $n=1$. Then $f:L\rightarrow L$ has least fixpoint $a_1=\mu x . f(x)$. Then $a_1=f(a_1)$.

    Suppose $n>1$. We fix a general $i$ such that $1<i<n$. The cases $i=1$ and $i=n$ are simplifications of the general case. First for each $j\neq i$, observe that $a_j$ is part of a vectorial fixpoint for $f$ specialised on $a_i$:
    \[Sp_{i,a_i}f(a_1,\dots,\widehat{a_i},\dots,a_n)=(a_1,\dots,\widehat{a_i},\dots,a_n)\]

    The above input and output vectors are of length $n-1$, and $Sp_{i,a_i} f$ is obviously monotone. Note that when $j>i$, $a_j$ is at the $(j-1)$-th coordinate of $Sp_{i,a_i}f$. By IH we have for each $j\neq i$
    \[a_j \geq \begin{cases}
        (Sp_{i,a_i}f)_j(\lambda k. ?k=j:a_j;nested(n-1,k,[j\coloneq a_j],Sp_{i,a_i}f)) & \text{if } j<i\\
        (Sp_{i,a_i}f)_{j-1}(\lambda k. ?k=j-1: a_j; nested(n-1,k,[j-1\coloneq a_j],Sp_{i,a_i}f)) & \text{if } j>i
    \end{cases} \]
    When $j<i$ we have by lemma \ref{lem_shift}
    \[nested(n-1,k,[j\coloneq a_j],Sp_{i,a_i}f) = \begin{cases}
        nested(n,k,[j\coloneq a_j, i\coloneq a_i],f) & \text{if } k<i \\
        nested(n,k+1, [j\coloneq a_j, i\coloneq a_i],f) & \text{if } k+1>i
    \end{cases} \] 
    
    When $j>i$ we have again by lemma \ref{lem_shift}
    \[nested(n-1,k,[j-1\coloneq a_j],Sp_{i,a_i}f) = \begin{cases}
        nested(n,k,[j\coloneq a_j, i\coloneq a_i],f) & \text{if } k<i \\
        nested(n,k+1, [j\coloneq a_j, i\coloneq a_i],f) & \text{if } k+1>i
    \end{cases} \] 

Let us fix some $j>i$. Then
\begin{multline*}
    a_j\geq (Sp_{i,a_i}f)_{j-1}(\lambda k. ?k=j-1: a_j; nested(n-1,k,[j-1\coloneq a_j],Sp_{i,a_i}f)) \\
    = f_j(\lambda h\in\{1,\dots,n\}. ?h=i:a_i; ?h=j:a_j; \\
        ?h<i: nested(n,h,[j\coloneq a_j, i\coloneq a_i],f); \\
        nested(n,h-1+1,[j\coloneq a_j,i\coloneq a_i],f) \\
    = f_j(\lambda h. ?h=i:a_i;?h=j:a_j;nested(n,h,[j\coloneq a_j,i\coloneq a_i],f))
\end{multline*}

By definition of least fixpoint we get
\begin{multline*}
    a_j\geq \mu {x_j}. f_j(\lambda h. ?h=i:a_i;?h=j:x_j;nested(n,h,[j\coloneq x_j,i\coloneq a_i],f)) \\
     = nested(n,j,[i\coloneq a_i],f)
\end{multline*}

Now fix some $j<i$. Then we have similarly
\begin{multline*}
    a_j\geq (Sp_{i,a_i}f)_{j}(\lambda k. ?k=j: a_j; nested(n-1,k,[j\coloneq a_j],Sp_{i,a_i}f)) \\
    = f_j(\lambda h\in\{1,\dots,n\}. ?h=i:a_i; ?h=j:a_j; \\
        ?h<i: nested(n,h,[j\coloneq a_j, i\coloneq a_i],f); \\
        nested(n,h-1+1,[j\coloneq a_j,i\coloneq a_i],f) \\
    = f_j(\lambda h. ?h=i:a_i;?h=j:a_j;nested(n,h,[j\coloneq a_j,i\coloneq a_i],f))
\end{multline*}

and we get the same inequality
\begin{equation*}
    a_j\geq nested(n,j,[i\coloneq a_i],f)
\end{equation*}

We substitute this inequality into the equation
\[a_i = f_i(\lambda j. j=i:a_i; a_j)\]
and we get
\begin{equation*}
    a_i \geq f_i(\lambda j. ?j=i:a_i;nested(n,j,[i\coloneq a_i],f))
\end{equation*}
which is exactly what is claimed.
\end{proof}

With Claim \ref{claim_ailarge} and definition of least fixpoint we immediately get half of what we are supposed to show:
\[a_i\geq \mu x_i.f_i(\lambda j. ?j=i:x_i; nested(n,j,[i\coloneq x_i],f) = nested(n,i,[],f)\]

For the other half, we write $a_i' = nested(n,i,[],f)$. Then we have
\[
    a_i' = f_i(\lambda j. ?j=i:a_i'; nested(n,j,[i\coloneq a_i'],f))
\]

\[Sp_{1,a_1'}f(x_1,\dots,x_{n-1}) = (f_2(a_1',x_1,\dots,x_{n-1}),\dots, f_n(a_1',x_1,\dots,x_{n-1}))\]

By IH, we have
\[\mu\begin{bmatrix}
    x_1 \\ \dots \\ x_{n-1}
\end{bmatrix}. \begin{bmatrix}
    (Sp_{1,a_1'}f)_1 \\ \dots \\ (Sp_{1,a_1'}f)_{n-1}
\end{bmatrix}
 = \begin{bmatrix}
     nested(n-1,1,[], Sp_{1,a_1'}f) \\ \dots \\ nested(n-1,n-1, [], Sp_{1,a_1'}f)
 \end{bmatrix}
\]
By lemma \ref{lem_shift}, the RHS equals 
\[\begin{bmatrix}
     nested(n,2,[1\coloneq a_1'], f) \\ \dots \\ nested(n,n,[1\coloneq a_1'], f)   
\end{bmatrix}\]
Write $a_i''= nested(n,i,[1\coloneq a_1'],f)$. Hence from the vectorial fixpoint for $Sp_{1,a_1'}f$ we have equations
\[\begin{cases}
    f_2(a_1',a_2'',\dots, a_n'') = a_1' \\
    f_3(a_1',a_2'',\dots, a_n'') = a_2'' \\
    \dots \\
    f_n(a_1',a_2'',\dots, a_n'') = a_n''
\end{cases}
\]
Moreover since $a_1' = nested(n,1,[],f) = \mu x_1.f_1(\lambda j. j=1:x_1; a_j'')$ we have
\[f_1(a_1',a_2'',\dots,a_n'') = a_1'\]

This implies that the vector $(a_1',a_2'',\dots,a_n'')$ is a fixpoint of $f$. Since $(a_1,\dots,a_n)$ is the least fixpoint, we get $a_1 \leq a_1'$. By a permutation argument we also get $a_i\leq a_i'$ for all $i>1$. This finishes the proof of theorem $\ref{thm_bekic}$.

\subsection{Concluding remarks}
We have presented an inductive proof of the $n$-ary Beki\v{c} principle. Our principal motivation for going into such technical detail is the need for this proof to be formalized in the proof assistant \verb|Isabelle|. Hence, one cannot merely present an intuitive argument that relies on exhibiting the explicit form of nested fixpoint for a number $n$ that is not fixed. Such an argument would require human imagination and the size of the argument grows in the order of $EXP(n)$. In this regard, induction is a higher-order reasoning mechanism that allows for a proof of fixed length to work for all numbers $n$. This is of course reflected in our proof as well.

\bibliographystyle{plain}
\bibliography{main}
\end{document}